\documentstyle[aps,pre,twocolumn,epsfig]{revtex}
\begin{document}
\twocolumn[\hsize\textwidth\columnwidth\hsize\csname@twocolumnfalse\endcsname
\title{Passive scalar turbulence in high dimensions}
\author{Andrea Mazzino$^{1,2}$ and Paolo Muratore-Ginanneschi$^{2}$ \\
\small{$^{1}$ INFM--Department of Physics, University of Genova, I--16146
Genova, Italy.}\\
\small$^2$ The Niels Bohr Institute, Blegdamsvej 17, DK-2100 Copenhagen, 
Denmark.}
\draft
\date{\today}
\maketitle
\begin{abstract}
Exploiting a Lagrangian strategy we present a
numerical study for both perturbative and
nonperturbative regions of the Kraichnan advection model. The major result
is the numerical assessment of the
first-order $1/d$-expansion 
by M. Chertkov, G. Falkovich, I. Kolokolov and V. Lebedev 
({\it Phys. Rev. E}, {\bf 52}, 4924 (1995))
for the fourth-order scalar structure function
in the limit of high dimensions $d$'s.
In addition to the perturbative
results, the behavior of the anomaly for the
sixth-order structure functions {\it vs} 
the velocity scaling exponent, $\xi$, is investigated and the
resulting behavior discussed.
\end{abstract}
]

A first principles
theory of turbulent advection was very recently developed in
the framework of the passive scalar model introduced by
R.H.~Kraichnan \cite{K68,K94} where the velocity field advecting the scalar
is Gaussian in space and white in time (i.e.~$\delta$-correlated). 
The main feature of such model is that its statistical description 
can be obtained from the solution of a set of closed linear partial
differential equations for the equal-time correlation functions 
\cite{K68}.\\
It was conjectured by Kraichnan \cite{K94} that, despite the closure of
the equations for the equal-time correlation functions, 
the system should display an {\it intermittent}
behavior. This means \cite{F95}
that the scaling exponents $\zeta_{2n}$ of the
$2n$-order scalar structure functions $S_{2n}$ are {\em anomalous}, namely
$
S_{2n} \equiv \langle [\theta({\bf r},t) - \theta({\bf 0},t)  ]^{2n}\rangle 
\propto r^{\zeta_{2n}}
$
with $\zeta_{2n} < n\zeta_{2} $ as $r$ goes to zero. This is in sharp
contrast with the {\it normal} scaling $\zeta_{2n}=n\zeta_2$ predicted 
on the basis of mean-field dimensional arguments.

Intermittent behavior was subsequently established by
Gaw\c{e}dzki and Kupiainen \cite{GK95},
Cher\-t\-kov {\it et al} \cite{CFKL95} and Shraiman and Siggia 
\cite{SS95,SS96}.
Anomalous scaling of the structure functions appears because 
of the presence of zero modes of the 
equations for the correlation 
functions \cite{GK95,CFKL95,BeGaKu,SS95,SS96,CF96}.\\ 
The second-order structure function was solved exactly 
in Ref.~\cite{K94} and it does not show any anomaly.  
Corrections to normal scaling start to appear for the fourth order
structure functions (the third, when scalar fluctuations are sustained 
by a large-scale gradient \cite{P96}). The anomalous exponents have been
up to now computed only perturbatively around limit cases where the 
structure functions of the Kraichnan model are known and display normal 
scaling.
The corresponding perturbative regions are\,: small $\xi$'s \cite{GK95,AA98}, 
large $d$'s \cite{CFKL95,CF96} and $\xi$ close to the Batchelor limit 
$\xi=2$ \cite{SS95,SS96,PSS97,BFL97}, 
$\xi$ being the scaling exponent of the advecting 
velocity field and $d$ the dimension of the space.
The first two expansions are regular, while for the
third one the relevant small parameter should be $\sqrt{2-\xi}$. This
is due to the preservation of the collinear geometry in the Batchelor
limit, leading to an angular non-uniformity in the perturbation
analysis \cite{BCKL95bis}.

The scaling behavior of high order structure functions
was investigated by means of instanton calculus \cite{FKLM96}.
The solutions obtained up to now  predict the saturation to a constant 
value of the scaling exponents $\zeta_n$'s as $n$ becomes large 
\cite{Y97,C97,BL98}.

It is of great interest both to develop efficient numerical strategies 
to test the predictions of the available perturbation theories -- 
and thus to give numerical confirmations of the
mechanisms responsible for the emergence of intermittency -- and to
investigate the nonperturbative regions, still not completely
accessible by analytical approaches.
Exploiting a new numerical strategy such program was pursued 
in Refs.~\cite{FMV98,FMNV99,CLMV00} 
(see also the review paper \cite{SS00})
where, in particular, the behavior of the anomaly
$\sigma_{2n}\equiv n\zeta_2-\zeta_{2n}$
was studied as a function of $\xi$ for $n=2$ both in two and
three dimensions.  The numerical experiments confirmed 
the perturbative predictions for $\xi\to0$ and $\xi\to2 $.
However, due to the difficulty of the numerical simulations no 
results were available for the behavior of $\sigma_{2n}$  as a 
function of $\xi$ for moments higher than the fourth and far from the 
two above perturbative regions.
A numerical study of the curve  $\sigma_{6}$ vs $\xi$ 
in the nonperturbative region of the Kraichnan model
is one of the aims of the present Letter. \\
The second aim is the investigation of the Kraichnan model
in large spatial dimensions. 
The $1/d$ expansion received solely a partial numerical confirmation
for the third-order structure function \cite{GPZ98}. The
value of $d$ investigated in Ref.~\cite{GPZ98} was not large enough 
to reach the perturbative regime for the fourth-order structure 
function and thus to verify the perturbative prediction of 
Ref.~\cite{CFKL95}. This is a second goal 
of the present Letter. Exploiting the Lagrangian strategy
presented in  Refs.~\cite{FMV98,FMNV99} we present two sets 
of numerical simulations which make possible the numerical assessment of the
$1/d$-expansion of Ref.~\cite{CFKL95}. We found that, for $\xi=0.8$,
the value of $d$ at which the perturbative regime studied in 
Ref.~\cite{CFKL95} takes place is $d\sim 30$.
The latter value reduces as
$\xi$ decreases, being of the order of $25$ for $\xi=0.6$.

Let us briefly recall the Kraichnan advection model \cite{K68,K94}.
In this model the velocity field ${\bf v}=\{v_\alpha,
\alpha=1,\ldots,d\}$ advecting the scalar is incompressible,
isotropic, Gaussian in space, white-noise in time; it has homogeneous
increments with power-law spatial correlations and a scaling exponent
$\xi$ in the range $0< \xi< 2$\,:
\begin{equation}
\left\langle [v_\alpha ({\bf r},t)-v_\alpha 
({\bf 0},0)][v_\beta ({\bf r},t)-v_\beta ({\bf 0},0)]
\right\rangle = 2 \delta (t) D_{\alpha\beta}({\bf r}),
\label{correlations}
\end{equation}
where,
\begin{eqnarray}
D_{\alpha\beta} ({\bf r})= D_0 r^{\xi}\left[\left(\xi+d-1\right)
\delta_{\alpha\beta}-\xi\,{r_{\alpha}
r_{\beta}\over r^2}\right].
\label{d-tensor}
\end{eqnarray}
The specific form inside the squared brackets is dictated by the 
incompressibility condition on ${\bf v}$.\\
The nonintermittent velocity field defined by (\ref{correlations})
advects a scalar field $\theta$
\begin{equation}
\partial_t\theta + {\bf v}\cdot {\bf \partial}\theta = \kappa\, 
\partial^2\theta + f
\label{FP}
\end{equation}
of which we want to investigate the statistical properties 
related to intermittency. Here $f$ 
is an  external forcing and $\kappa$ is the molecular diffusivity. \\
The forcing term permits to attain a stationary state defined by
the balance between production of scalar variance
(related to $f$) and its dissipation (related to $\kappa\partial^2\theta$). 
The net result is that $\langle \theta^2\rangle$ 
is finite in the steady state.\\
We shall assume  the (random) forcing $f$  of zero mean,
isotropic, Gaussian, white-noise in time and homogeneous. Its
correlation is specified by\,:
\begin{equation}
\langle f({\bf r},t)\,f({\bf 0},0)\rangle =
\chi\left(r/L\right)\,\delta(t),
\label{fcorr}
\end{equation}
with $\chi(0)>0$ and $\chi\left(r/L\right)$ a rapidly decreasing function  
for $r\gg L$ where $L$ is the (forcing) integral scale.

\begin{figure}
\begin{center}
\mbox{\hspace{0.0cm}\psfig{file=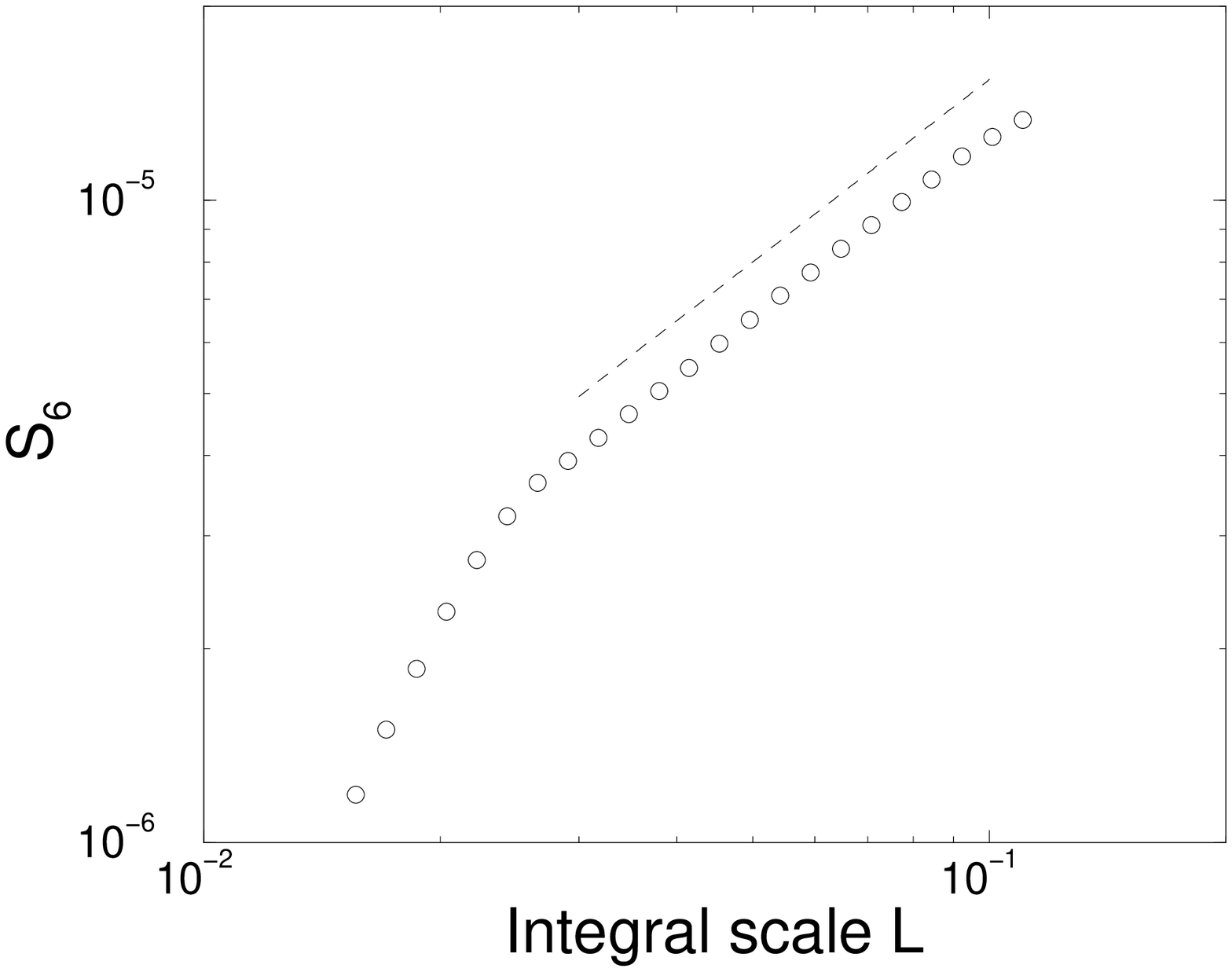,height=7cm,width=7cm}}
\end{center}
\vspace{-1.5cm}
\begin{center}
\mbox{\hspace{0.0cm}\psfig{file=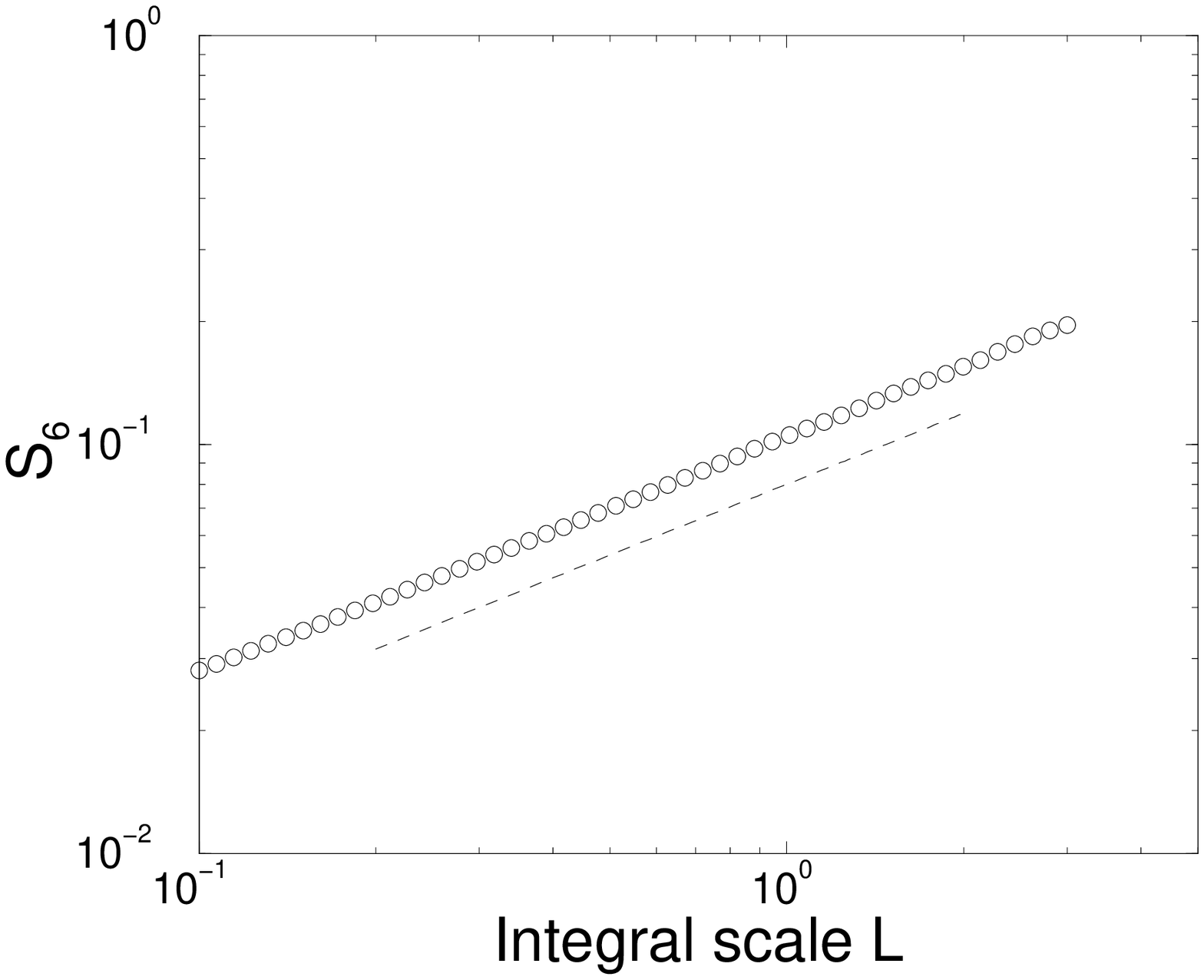,height=7cm,width=7cm}}
\end{center}
\vspace{0.0cm}
\caption{Three-dimensional sixth-order structure functions
$S_6$ {\it vs} $L$ for $\xi=0.9$ (above) and $\xi=1.5$ (below). 
The dashed straight lines are the best fit slopes calculated in the
scaling region. Their values give the anomalies reported in 
Fig.~\protect\ref{anomalie} (upper curve). Number of realizations
$\sim 10^7$.}
\label{fig1}
\end{figure}

\begin{figure}
\begin{center}
\mbox{\hspace{0.0cm}\psfig{file=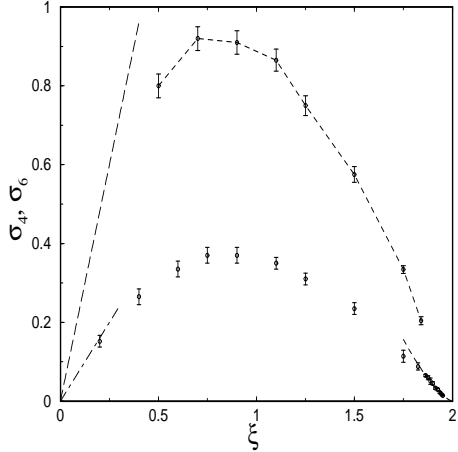,height=7cm,width=7cm}}
\end{center}
\vspace{-0.5cm}
\caption{The anomaly $\sigma_6\equiv 3\zeta_2-\zeta_6$ 
for the sixth-order structure
functions in  three dimensions (upper curve) and, for comparison,
the $\sigma_4$ curve already published in 
Ref.~\protect\cite{FMV98} (lower curve). The dashed straight line
on the left is the first-order prediction $\sigma_6=12\,\xi/5$ by
Bernard {\it et al} \protect\cite{BeGaKu}. Notice the shift on
the left of the maximum of the $\sigma_6$ curve  with respect 
to the maximum of $\sigma_4$.}
\label{anomalie}
\end{figure}

Rather than integrating the partial stochastic differential equation 
(\ref{FP})
(that is a prohibitive task already in two dimensions due to the 
$\delta$-correlation of ${\bf v}$), 
we exploit the Lagrangian formulation of the passive scalar 
dynamics as in Refs.~\cite{FMV98,FMNV99} in order
to simulate particle trajectories  by Monte--Carlo
methods.\\
We recall that when such method is adopted the evaluation of 
structure functions is reduced to the study of the 
statistical properties of the random variable describing the time
spent by pairs of particle with mutual distance less than
the integral scale $L$.  Generally, the distance between pairs of
particles tends to increase with the elapsed time but, occasionally,
particles may come very close and stay so; the phenomenon is the source 
of scaling anomalies.
 The main advantage of the Lagrangian strategy is that one does not have 
to generate the whole velocity field, but just the velocity field over
the Lagrangian trajectories. Simulations in very high space dimensions
as well as the investigation of high-order moments
become thus feasible.\\
 We now present results for structure functions up to sixth order.
The three-dimensional results relative to the fourth-order structure 
functions have been  already published in
Ref.~\cite{FMV98}. They will be reported in the following  
for comparison with our new data.\\
The $L$-dependence of $S_6(r;L)$ is shown for the three-dimensional case
in Fig.~\ref{fig1} for $\xi=0.9$ (above) and $\xi=1.5$ (below). 
Similar plots have been obtained also for other values of $\xi$.
In both cases the scaling region is indicated by a dashed
straight line the slope of which yields the anomaly 
$\sigma_6\equiv 3\zeta_2-\zeta_6$. 
 The measured slopes for $\xi=0.9$ and $\xi =1.5$
 are $\sigma_6=0.91\pm 0.03$ and $\sigma_6=0.58\pm 0.02$, respectively.
The error bars are obtained
by analyzing the fluctuations of local scaling exponents over smaller
ratios of values for $L$.
As explained in Ref.~\cite{FMNV99} the number of realizations needed 
to obtain high quality scaling as the ones displayed in~\cite{FMV98} 
increases rapidly with decreasing $\xi$.
This is the reason why we restricted our analysis to 
the interval $\xi \geq 0.5$.\\
The results for the anomalies {\it vs} $\xi$ are summarized in
Fig.~\ref{anomalie}. The upper curve refers to $\sigma_6$, lower curve to
 $\sigma_4$ (the latter has been already published in Ref.~\cite{FMV98}).\\
Some remarks on the behavior of $\sigma_6$
are worth. First, we note that when $\xi$ decreases from 2 to
0 the anomaly $\sigma_6$ grows at first, achieves a maximum and 
finally decreases as in the case of $\sigma_4$.  

We have evidence that $\sigma_6$ tends to
vanish for $\xi\to 0$ as follows from the perturbative predictions 
\cite{AA98,BeGaKu} the leading order of which $\sigma_6 
= 12\,\xi/5$ is shown as a dashed straight line on the left of 
Fig.~\ref{anomalie}.
The fact that the anomaly for $S_6$ is higher than the one for $S_4$
is an immediate consequence of H\"{o}lder inequalities \cite{F95}.

More interesting is the fact that the maximum of the anomaly occurs for a value
of $\xi$ which for $S_6$ is smaller than for $S_4$. 
This can be explained as follows. Near $\xi =0$ the dynamics is dominated by
the nearly ultraviolet-divergent eddy diffusion. The latter arises
from small-scale fluctuations averaged over time scales much larger
than the typical ones. Large fluctuations are thus averaged out
and appear strongly depleted.
On the contrary, near $\xi=2$ the dynamics
is dominated by the nearly infrared-divergent stretching. 
No averaging of the large fluctuations now
takes place and the net result is that they strongly contribute
to the scalar statistics. Since higher moments
are more sensitive to large fluctuations than the lower ones, 
one expects  stretching to dominate the dynamics for wider
ranges of $\xi$ as the order $n$ of the moment increases.
As a consequence the value of $\xi$ where the effects of stretching 
and diffusion balance (i.e. where the maximum of the anomaly is expected) 
should move toward the left as $n$ increases in agreement with the 
results reported in Fig.~\ref{anomalie}.\\
Let us now turn to the behavior of fourth-order structure
functions for large space dimensions. In such limit the first-order 
perturbative prediction in $1/d$ gives for the 
fourth-order structure functions the anomaly $4\xi/d$ 
\cite{CFKL95,CF96}. The Lagrangian strategy allows to verify
numerically 
the perturbative prediction.\\
\begin{figure}
\begin{center}
\mbox{\hspace{0.0cm}\psfig{file=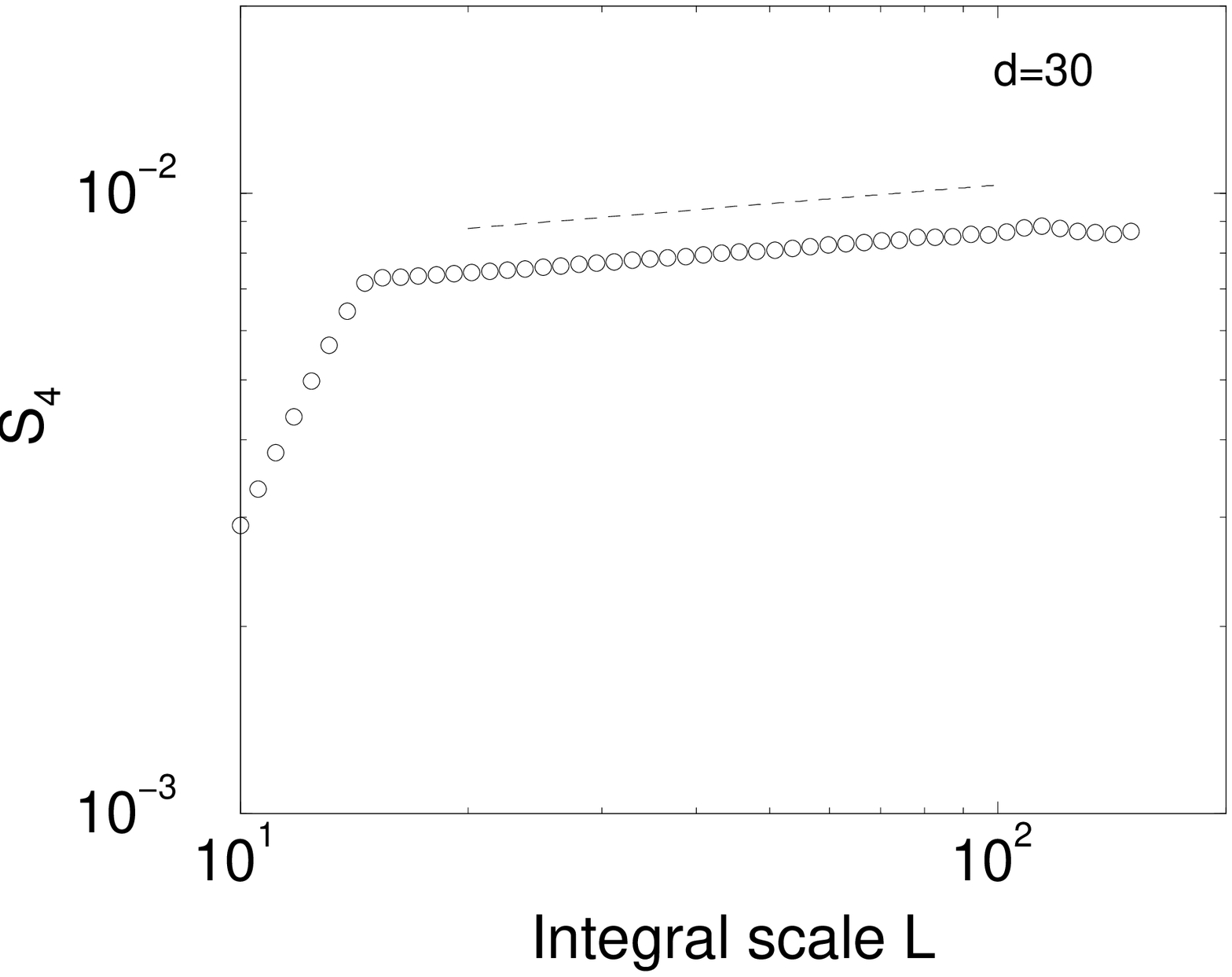,height=5.5cm,width=7cm}}
\end{center}
\vspace{-1.cm}
\begin{center}
\mbox{\hspace{0.0cm}\psfig{file=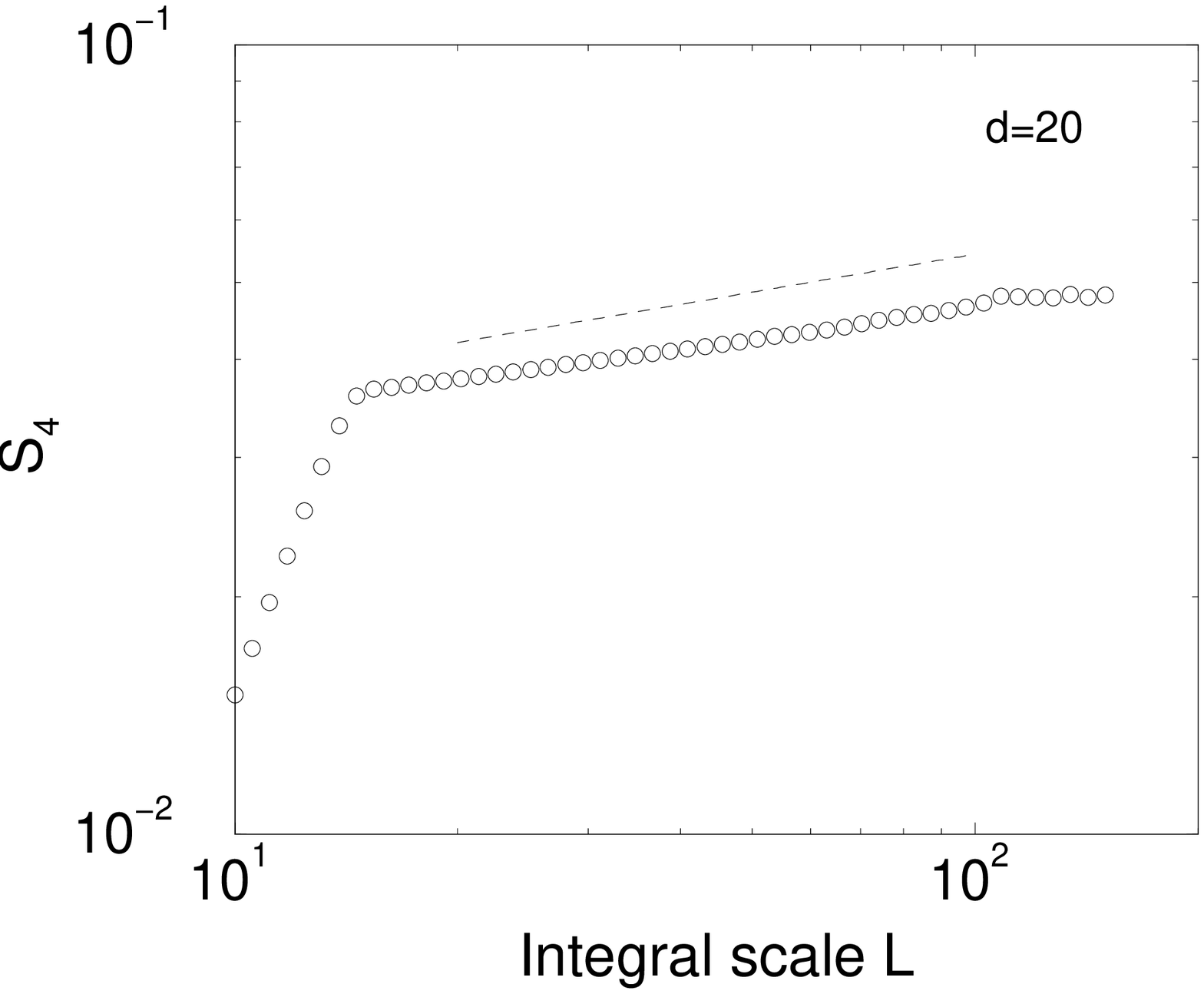,height=5.5cm,width=7cm}}
\end{center}
\vspace{-1.cm}
\begin{center}
\mbox{\hspace{0.0cm}\psfig{file=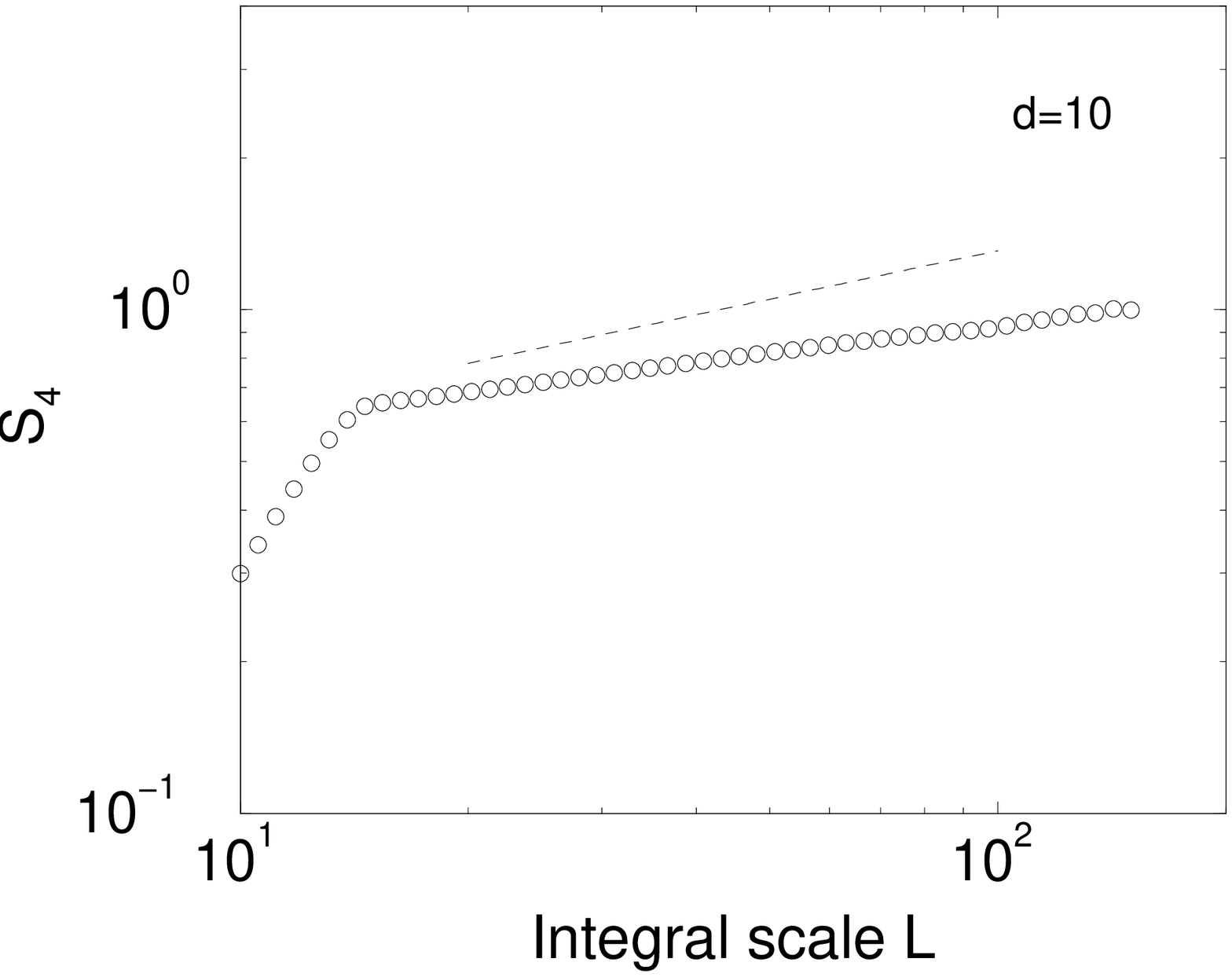,height=5.5cm,width=7cm}}
\end{center}
\vspace{-1.cm}
\begin{center}
\mbox{\hspace{0.0cm}\psfig{file=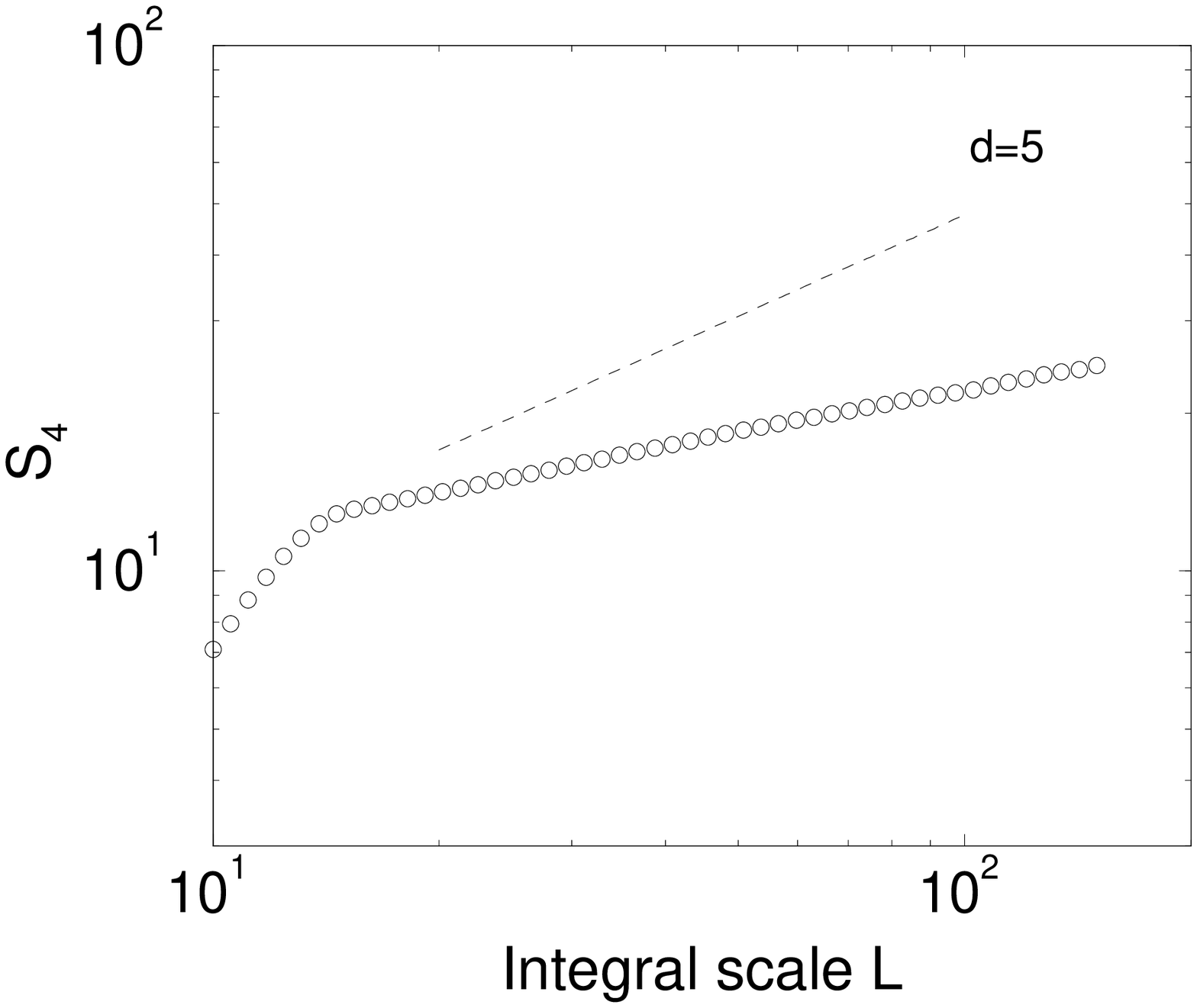,height=5.5cm,width=7cm}}
\end{center}
\vspace{0.0cm}
\caption{Fourth-order structure functions for $\xi=0.8$ and $d$ ranging
from 5 to 30. Dashed straight lines represent the anomaly from the
$1/d$-expansion by Chertkov {\it et al} \protect\cite{CFKL95}. Number of
realizations $\sim 10^6$.}
\label{ddimxi0.8}
\end{figure}

\begin{figure}
\begin{center}
\mbox{\hspace{0.0cm}\psfig{file=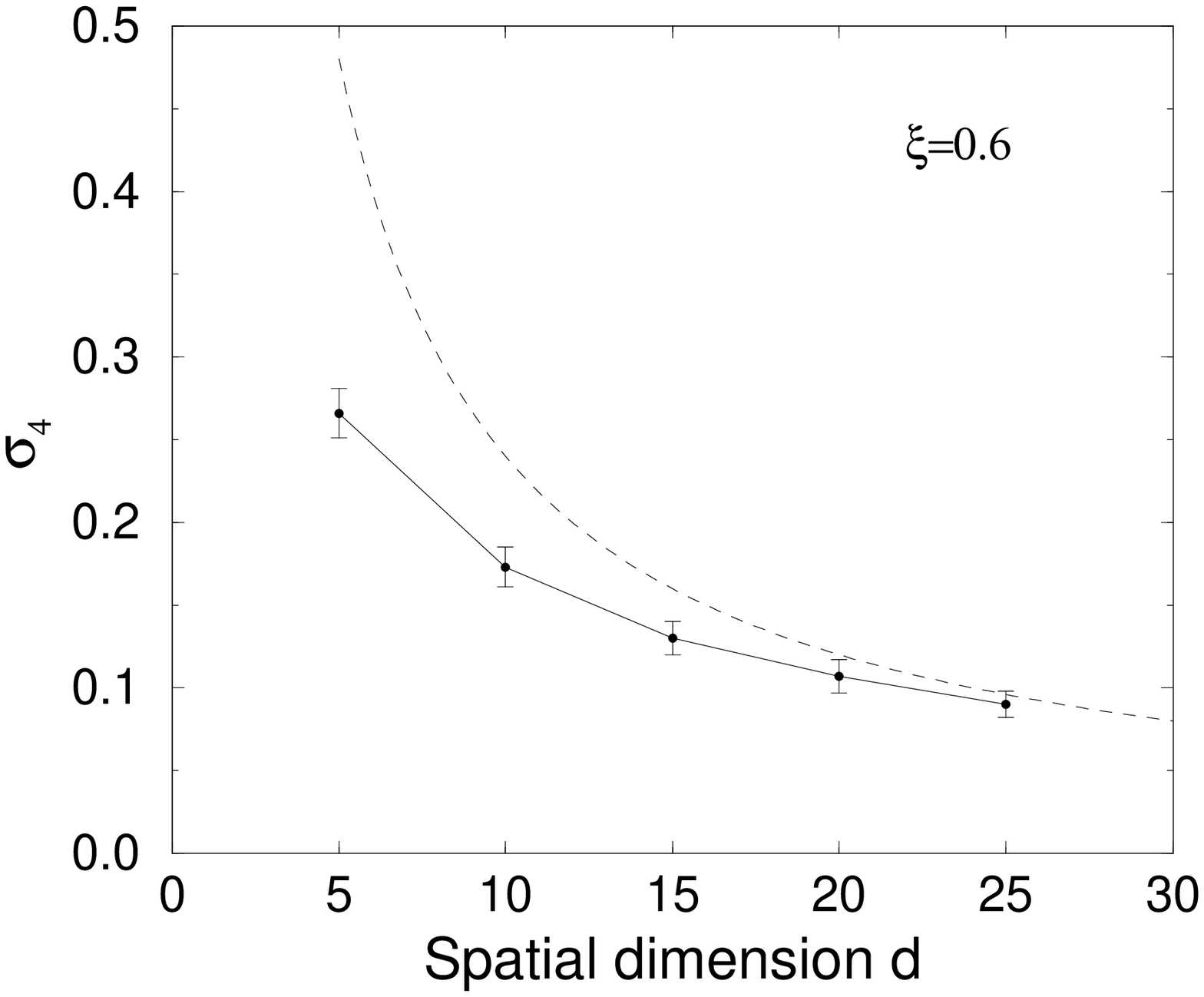,height=6cm,width=7cm}}
\end{center}
\vspace{-1.cm}
\begin{center}
\mbox{\hspace{0.0cm}\psfig{file=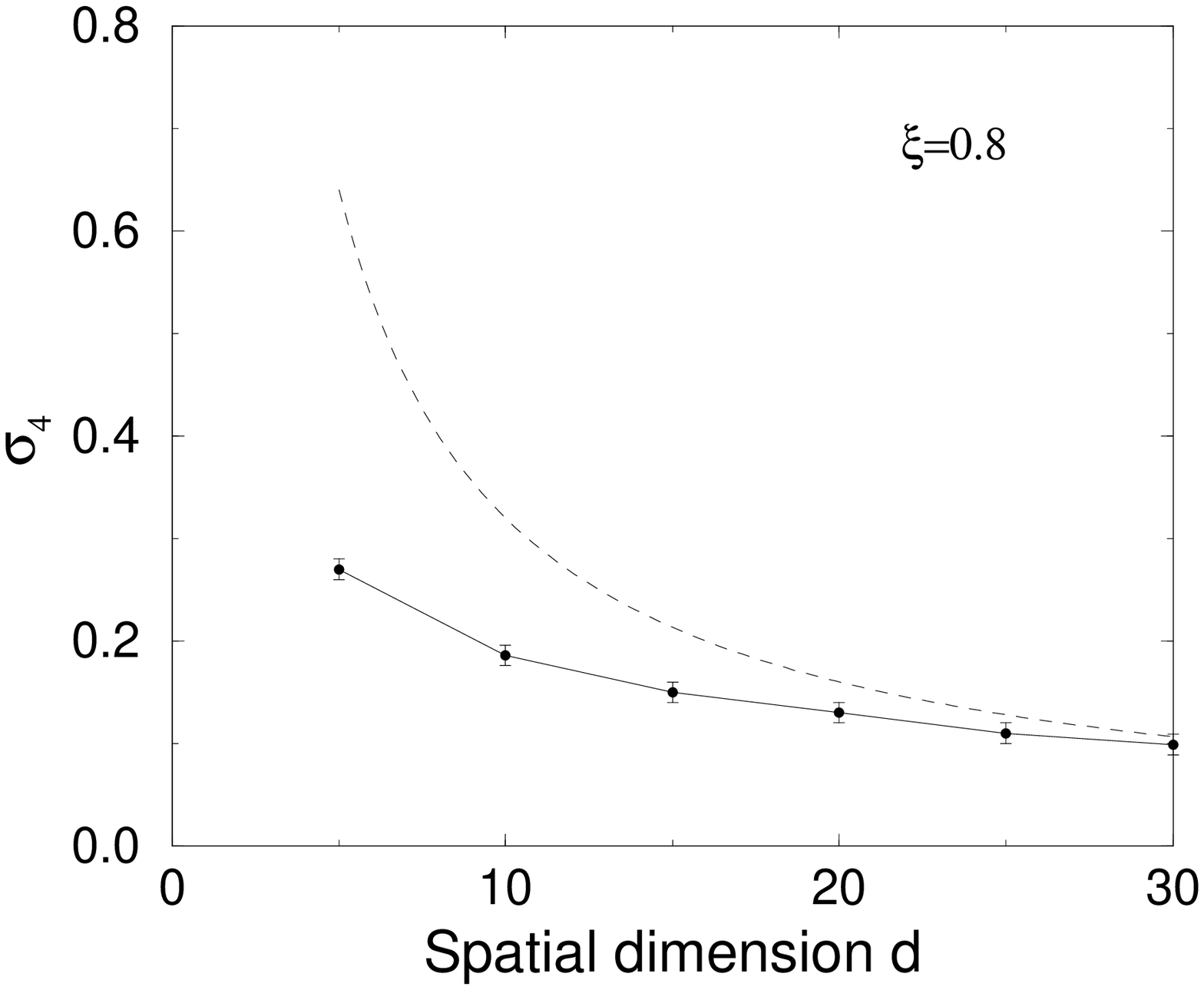,height=6cm,width=7cm}}
\end{center}
\vspace{-0.5cm}
\caption{The anomalies $\sigma_4$ {\it vs}
$d$ for $\xi=0.6$ and $\xi=0.8$.
Dashed lines are the anomalies $4\xi/d$ from the $O(1/d)$ perturbation
theory of Ref.~\protect\cite{CFKL95}.}
\label{riass}
\end{figure}

 We performed two sets of simulations for $\xi=0.6$ and $\xi=0.8$
and different spatial dimensions, $d$, from 5 to 30. 
The fourth-order structure functions
for  $d=5$, $10$, $20$ and $30$ are shown in
Fig.~\ref{ddimxi0.8} for $\xi=0.8$. Similar scaling laws have been
observed also for $\xi=0.6$. 
Dashed straight lines represent the slopes
$4\xi/d$.
For $d=30$ the anomaly obtained from numerical
simulations (i.e. the slopes of the curve with circles) is
practically indistinguishable  from the perturbative expression. 
The discrepancy increases rapidly as $d$ reduces. The results 
are summarized in Fig.~\ref{riass}, where the anomaly $\sigma_4$
is shown as a function of $d$ for $\xi=0.6$
and $\xi=0.8$. The fact that for $\xi=0.6$
the perturbative regime starts for a value of $d$ 
($\sim 25$) lower than the one at $\xi=0.8$ ($d\sim 30$) is the
consequence of the fact that the small parameter in the perturbation
theory is $\propto 1/[d (2-\xi)]$ rather than $1/d$. 

In conclusion, we verified the $1/d$-expansion for 
the fourth-order structure functions.
We presented two set of simulations corresponding to two different 
values of the velocity scaling exponent $\xi$.
 For $\xi=0.8$ the perturbative regime sets in for $d\sim 30$. 
This value reduces at $d\sim 25$ for $\xi=0.6$. The result is expected 
as $1/[d (2-\xi)]$ is the relevant small parameter in the 
perturbation theory.
We also studied the behavior of the anomaly for the
sixth-order structure functions vs the velocity scaling exponent
$\xi$. We identified and discussed two competing mechanisms 
which control the position of the maximum of the anomaly
along the $\xi$-axis.

We are grateful to A.~Celani, G.~Falkovich and M.~Vergassola 
for discussions and suggestions. 
AM was partially supported by the INFM project GEPAIGG01.
PMG was supported by the Europeam grant ERB4001GT962476.
Part of this work has been done during the research program 
{\em Physics of Hydrodynamic Turbulence} at the
Institute for Theoretical Physics of the University of California, S. Barbara.
Simulations were performed within the INFM Parallel Computing Initiative.


\begin{thebibliography}{99}

\bibitem{K68}
R.H.~Kraichnan, {\it Phys. Fluids}, {\bf 11}, 945, (1968).

\bibitem{K94}  R.H.~Kraichnan, {\it Phys. Rev. Lett.}, {\bf 52}, 1016, (1994).

\bibitem{F95} U.~Frisch, {\it Turbulence. The Legacy of
A.N.~Kolmogorov}, Cambridge Univ. Press, Cambridge, (1995).

\bibitem{GK95} K.~Gaw\c{e}dzki and A.~Kupiainen, 
{\it Phys. Rev.  Lett.}, {\bf 75}, 3834, (1995).

\bibitem{CFKL95}
M.~Chertkov, G.~Falkovich,
I.~Kolokolov and V.~Lebedev, {\it Phys. Rev. E}, {\bf 52}, 4924 (1995).

\bibitem{SS95}
B.I.~Shraiman and E.D.~Siggia, {\it C.R. Acad. Sci.}, {\bf 321}, S\'erie II,
279, (1995).

\bibitem{SS96}
B.I.~Shraiman and E.D.~Siggia, {\it Phys. Rev. Lett.}, {\bf 77}, 2463, (1996).

\bibitem{AA98} L.Ts.~Adzhemyan and N.V. Antonov,
{\it Phys. Rev. E}, {\bf 58}, 7381 (1998). 

\bibitem{BeGaKu}
D.~Bernard, K.~Gaw\c{e}dzki and A.~Kupiainen, Phys. Rev. E {\bf 54}, 2564 
(1996).

\bibitem{CF96}
M.~Chertkov and G.~Falkovich, 
{\it Phys. Rev. Lett}, {\bf 76}, 2706 (1996).

\bibitem{P96} A. Pumir, {\it Europhys. Lett.}, {\bf 34}, 25 (1996).


\bibitem{PSS97}
A.~Pumir, B.I.~Shraiman and E.D.~Siggia,
{\it Phys. Rev. E}, {\bf 55}, R1263 (1997).

\bibitem{BFL97}
E.~Balkovsky, G.~Falkovich and V.~Lebedev,
{\it Phys. Rev. E}, {\bf 55}, R4881 (1997).


\bibitem{BCKL95bis}
E.~Balkovsky, M.~Chertkov, I.~Kolokolov and V.~Lebedev,
{\it JETP Lett.}, {\bf 61}, 1049 (1995).

\bibitem{FKLM96}
G.~Falkovich, I.~Kolokolov, V.~Lebedev and A.~Migdal,
{\it Phys. Rev. E}, {\bf 54}, 4896, (1996).

\bibitem{Y97}
V.~Yakhot,
{\it Phys. Rev. E}, {\bf 55}, 329 (1997).

\bibitem{C97}
M.~Chertkov,
{\it Phys. Rev. E}, {\bf 55}, 2722 (1997).

\bibitem{BL98}
E.~Balkovsky and V.~Lebedev,
{\it Phys. Rev. E},  {\bf 58}, 5776 (1998).

\bibitem{FMV98}
U.~Frisch, A.~Mazzino,  and M.~Vergassola,
{\it Phys. Rev. Lett.}, {\bf 80}, 5532 (1998).

\bibitem{FMNV99}
U.~Frisch, A.~Mazzino,  A.~Noullez, and M.~Vergassola,
{\it Phys. Fluids}, {\bf 11}, 2178 (1999).

\bibitem{CLMV00}
A.\ Celani, A.\ Lanotte, A.\ Mazzino and M.\ Vergassola,
{\it Phys. Rev. Lett.}, {\bf 84},  2385 (2000).

\bibitem{SS00}
B.I.~Shraiman and E.D.~Siggia, 
{\it Nature}, {\bf 405}, 639 (2000).


\bibitem{GPZ98}
O.~Gat, I.~Procaccia and R.~Zeitak, {\it Phys. Rev. Lett.}, {\bf 80}, 
5536, (1998).

\end{thebibliography}
\end{document}